\documentclass[%
 reprint, 
superscriptaddress,
 amsmath,amssymb,
prc,
]{revtex4-1}

\usepackage{graphicx}
\usepackage{dcolumn}
\usepackage{bm}

\begin{document}

\title{Muonic X-Ray Measurement for the Nuclear Charge Distribution:\\ the Case of Stable Palladium Isotopes}

\author{T. Y. Saito$^*$}
 \affiliation{Graduate School of Science, the University of Tokyo, 7-3-1 Hongo, Bunkyo-ku, 113-0033 Tokyo, Japan}
 \email{saito@cns.s.u-tokyo.ac.jp}
\author{M. Niikura}%
 \affiliation{Graduate School of Science, the University of Tokyo, 7-3-1 Hongo, Bunkyo-ku, 113-0033 Tokyo, Japan}
\author{T. Matsuzaki}
 \affiliation{RIKEN Nishina Center, RIKEN, 2-1 Hirosawa, Wako-shi, 351-0198 Saitama, Japan}
\author{H. Sakurai}
 \affiliation{Graduate School of Science, the University of Tokyo, 7-3-1 Hongo, Bunkyo-ku, 113-0033 Tokyo, Japan}
 \affiliation{RIKEN Nishina Center, RIKEN, 2-1 Hirosawa, Wako-shi, 351-0198 Saitama, Japan}
\author{M.~Igashira}
 \affiliation{Laboratory for Advanced Nuclear Energy, Tokyo Institute of Technology, 2-12-1 Ookayama, Meguro-ku, 152-8550 Tokyo, Japan}
\author{H.~Imao}
 \affiliation{RIKEN Nishina Center, RIKEN, 2-1 Hirosawa, Wako-shi, 351-0198 Saitama, Japan}
\author{K.~Ishida} 
 \affiliation{RIKEN Nishina Center, RIKEN, 2-1 Hirosawa, Wako-shi, 351-0198 Saitama, Japan}
 \author{T.~Katabuchi}
\affiliation{Laboratory for Advanced Nuclear Energy, Tokyo Institute of Technology, 2-12-1 Ookayama, Meguro-ku, 152-8550 Tokyo, Japan}
\author{Y.~Kawashima}
 \affiliation{Research Center for Nuclear Physics, Osaka University, 10-1 Mihogaoka, Ibaraki-shi, 567-0047 Osaka, Japan}
\author{M.~K.~Kubo}
 \affiliation{Graduate School of Science, International Christian University, 3-10-2 Osawa, Mitaka-shi, 181-0015 Tokyo, Japan}
\author{Y.~Miyake}
 \affiliation{Institute of Materials Structure Science, High Energy Accelerator Research Organization, 1-1 Oho, Tsukuba-shi, 305-0801 Ibaraki, Japan}
\author{Y.~Mori}
 \affiliation{Institute for Integrated Radiation and Nuclear Science, Kyoto University, 2 Asashiro-Nishi, Kumatori-cho, Sennan-gun, 590-0494 Osaka, Japan}
\author{K.~Ninomiya}
 \affiliation{Graduate School of Science, Osaka University, 1-1 Machikaneyama-cho, Toyonaka-shi, 560-0043 Osaka, Japan}
\author{A.~Sato}
 \affiliation{Graduate School of Science, Osaka University, 1-1 Machikaneyama-cho, Toyonaka-shi, 560-0043 Osaka, Japan}
 \affiliation{Research Center for Nuclear Physics, Osaka University, 10-1 Mihogaoka, Ibaraki-shi, 567-0047 Osaka, Japan}
\author{K.~Shimomura}
 \affiliation{Institute of Materials Structure Science, High Energy Accelerator Research Organization, 1-1 Oho, Tsukuba-shi, 305-0801 Ibaraki, Japan}
\author{P.~Strasser}
 \affiliation{Institute of Materials Structure Science, High Energy Accelerator Research Organization, 1-1 Oho, Tsukuba-shi, 305-0801 Ibaraki, Japan}
\author{A.~Taniguchi}
\affiliation{Institute for Integrated Radiation and Nuclear Science, Kyoto University, 2 Asashiro-Nishi, Kumatori-cho, Sennan-gun, 590-0494 Osaka, Japan}
\author{D.~Tomono}
 \affiliation{Research Center for Nuclear Physics, Osaka University, 10-1 Mihogaoka, Ibaraki-shi, 567-0047 Osaka, Japan}
\author{Y.~Watanabe}
 \affiliation{RIKEN Nishina Center, RIKEN, 2-1 Hirosawa, Wako-shi, 351-0198 Saitama, Japan}

\date{\today}

\begin{abstract}

\begin{description}
\item[Background]
The nuclear charge radi1us and distribution are the most fundamental quantities of the atomic nucleus.
From the muonic transition energies, the absolute charge radius has been experimentally obtained, while there have been no established methods to discuss the distribution.
\item[Purpose]
The muonic transition energies for five palladium isotopes with a mass number $A = 104$, $105$, $106$, $108$, and $110$ were measured.
The procedure to deduce the charge radii and the method to discuss the charge distribution from the muonic transition energies are proposed.
\item[Method]
The experiment was performed at the MuSIC-M1 beamline at Research Center for Nuclear Physics, Osaka University.
A continuous muon beam impinged on the enriched palladium targets.
Muonic X rays were measured by high-purity germanium detectors.
\item[Results]
The muonic transition energies up to $4f$-$3d$ transitions were determined for five palladium isotopes.
\item[Discussion and conclusion]
The root-mean-square charge radii are deduced assuming the two-parameter Fermi distribution.
The charge distribution of the nucleus is discussed by employing the Barrett model.
The muonic transition energies of the $3d$-$2p$ transitions are crucial for discussing both the charge radius and the charge distribution.
\end{description}
\end{abstract}

\maketitle


\section{Introduction}
The nuclear charge radius is one of the most fundamental quantities of the atomic nucleus~\cite{Fricke1995,Angeli2013} since it is affected by the nuclear structure, such as the nuclear deformation and the nucleon pairing of the ground state. 
The absolute values of the charge radius are experimentally deduced from the electron scattering cross-sections and the muonic X-ray transition energies for stable nuclei, while the relative values are extracted from optical isotope shift measurements, which can be extended to unstable nuclei.
In the analysis of the electron scattering experiments, the charge distribution is deduced from the Fourier transformation of the measured form factor, and the radius, usually the root-mean-square (rms) radius, is a simple contracted form of the charge distribution.
The charge distributions have been systematically measured for almost all stable nuclei. Recently, the electron scattering method has also been applied to unstable nuclei~\cite{Wakasugi2013,Suda2017,Tsukada2017}.
Similarly, proton scattering experiments provide the nucleon density distribution~\cite{Terashima2008,Zenihiro2010}.
These distributions are important for understanding further details of the nuclear structure, such as shell evolution and neutron skin thickness.




The nuclear charge radii have also been deduced from the X-ray transition energies of a muonic atom. 
The muon has about 200~times larger mass than the electron and the atomic radius of the muonic atom is much smaller than that of the ordinary electronic atom.
The binding energy of the muonic atom is sensitive to the charge distribution of the nucleus.
In the first approximation, the muonic atom can be described as a two-body system of the muon and the nucleus.
The theoretical interpretation of the muonic atom is much simpler than that of the ordinary electronic atom, in which many atomic electrons interact with each other.
Therefore, one can determine the absolute nuclear charge radius by the muonic X-ray measurement.


Contrary to the simple description of the muonic atom, one has to introduce a model for the charge distribution to deduce the charge radii from the muonic X-ray energies.
The transition energies cannot be translated directly to the charge distribution, unlike the Fourier transformation of the form factor of the electron scattering.
This translation problem is usually treated by assuming a two-parameter Fermi (2pF) distribution with fixed surface diffuseness as the charge distribution~\cite{Fricke1995, Angeli2013}. 
This extreme simplification leads to an uncertainty of the model on the charge radii, which has not been extensively evaluated.


The higher-order transitions of the muonic atom are the key to reducing the uncertainties on the charge radii.
The number of experimental inputs limits the number of model parameters.
By including higher transitions, the surface diffuseness in the 2pF charge distribution, which is fixed in the standard treatment mentioned above, can also be experimentally determined.
The pioneering study by Bergem \textit{et al.}.~achieved 0.02$\%$ precision for the rms radius of $\mathrm{^{208}Pb}$, in which higher-order X rays up to the $4f$-$3d$ transitions were measured~\cite{Bergem1988}. 
Hence, the measurement of the higher-order X-ray transitions is essential for the precise determination of the charge radii.



The higher-order transitions also contain information on the charge distribution, and can be used as a benchmark of results obtained from electron scattering and theoretical calculations.
As one of the theoretical approaches to interpret the transition energies of the muonic atoms, Barrett proposed an empirical approximation, which provides the method to discuss the consistency among the muonic transition energies and the electron scattering~\cite{Barrett1970}.
In this paper, we propose employing Barrett's approach using several transitions to discuss the charge distribution from the muonic transition energies.


Recent technical progress demonstrated the possibility of forming the muonic atom with unstable nuclei~\cite{Strasser2005,Strasser2009}.
The muonic X rays of unstable nuclei are particularly of interest since the exotic behavior of the charge distribution among the unstable nuclei can be precisely and directly investigated by the muonic X-ray measurement.
Such exotic structures of unstable nuclei can only be investigated by including the higher transitions.
Therefore the establishment of interpretation methods for the higher muonic transitions is strongly required.


In this paper, we experimentally obtained the muonic X-ray transition energies of $\mathrm{^{104, 105, 106, 108, 110} Pd}$ including higher transitions and discuss how the higher transitions improve the determination of the charge distribution parameters.
Previous results of the muonic X-ray measurements are summarized in several compilation tables~\cite{Engfer1974,Fricke1995,Angeli2013}.
These compilations contain muonic X-ray energies and the nuclear charge parameters such as rms radii and Barret moments.
The rms radii are determined from only the lowest $2p$-$1s$ transitions in these compilations.
However, for some nuclei, such as stable palladium isotopes, which we investigated in this work, the experimental results were not published as original papers~\cite{Hack1989}.
This situation could undermine the fundamental importance of the nuclear charge parameters in natural science.

In addition to the X-ray energies, nuclear muon capture, one of the decay paths of the muonic atom, is of interest.
The muon bound by the nucleus, with the proton number $Z$ and the mass number $A$, decays via two weak processes, \textit{i.e.}~decay to an electron as same as a muon in a vacuum ($\mu$-$e$ decay): $\mu^- \to e^- + \bar{\nu_e} + \nu_{\mu}$ and nuclear muon capture: $(Z,A) + \mu^- \to (Z-1,A)^* + \nu_{\mu}$~\citep{Measday2001}.
The former process is dominant in the muonic atom of light nuclei, and the lifetime is close to that of the muon in a vacuum, namely $2.2$~$\mathrm{\mu}$s.
The latter process becomes more dominant, and the lifetime becomes shorter as $Z$ increases.
The $Z$ dependence of the total muon capture rate $\Lambda_C$ of the muonic atoms of heavier nuclei generally follows the phenomenological $Z^4$-dependence so-called Primakoff rule~\citep{Primakoff1959}.
The precise measurementa, however, have revealed that $\Lambda_C$ depends on the structure of the excited states of a produced nucleus and the nuclear matrix elements between the ground state of a target nucleus and the excited states of the produced nucleus.
Since the nuclear muon capture is a weak process with a high momentum transfer, it is similar to the virtual transition of a neutrinoless double $\beta$~decay ($0\nu \beta \beta $).
The measured muon capture rate thus constrains the theoretical calculation of the $0\nu \beta \beta$ matrix elements~\citep{Kolbe2000,Zinner2006,Marketin2009}.
In this work, we also determined the lifetime and corresponding $\Lambda_C$ of the palladium isotopes.

The rest of this paper is organized as follows. In Sec.~\ref{sec_exp}, the experimental setup and analysis procedure will be presented. The total muon capture rate will be discussed in Sec.~\ref{sec_lambdac}.
The nuclear charge radii for the Pd isotopes will be deduced in Sec.~\ref{sec_radius}, and the charge distribution will be discussed introducing the Barrett model in Sec.~\ref{sec_barret}.
We conclude this work in Sec.~\ref{sec_conclusion}.


\section{Experiment and analysis}\label{sec_exp}

\begin{table*}[]
  \begin{center}
    \caption{The composition of each palladium enriched target. The targets with * symbols are metal powder and the others are metal disks~\cite{Terada2015}. }
      \begin{tabular}{p{4em}p{5em}p{5em}p{5.5em}p{5.5em}p{5.5em}p{5.5em}p{5.5em}p{5.5em}} \hline \hline
    target & chemical purity [\%] & weight [g]  & $^{102}{\rm Pd}$ & $^{104}{\rm Pd}$ & $^{105}{\rm Pd}$ &$^{106}{\rm Pd}$ & $^{108}{\rm Pd}$ & $^{110}{\rm Pd}$  \\ \hline
    $^{104}{\rm Pd}$     & 99.97 & 0.69  & $<$0.02  & 98.4(1)  & 1.05(5) & 0.35(3)  & 0.18(2)   & $<$0.05  \\
    $^{105}{\rm Pd}^{*}$ & 99.97 & 0.49  & 0.033(6) & 0.236(4) & 97.9(7) & 1.407(8) & 0.311(4)  & 0.112(2) \\
    $^{106}{\rm Pd}^{*}$ & 99.97 & 0.85  & $<$0.03  & 0.06     & 0.68    & 98.4(2)  & 0.8       & 0.06     \\
    $^{108}{\rm Pd}$     & 99.97 & 0.657 & $<$0.02  & 4.8(1)   & 0.15(3) & 0.90(5)  & 93.80(15) & 0.30(3)  \\
    $^{110}{\rm Pd}$     & 99.99 & 0.69  & $<$0.05  & 0.1      & 0.35    & 0.5      & 0.7       & 98.3(2)  \\
    \hline \hline
      \end{tabular}
    \label{tab:target}
  \end{center}
\end{table*}

The experiment was performed at Research Center for Nuclear Physics (RCNP) in Osaka University, Japan.  A continuous muon beam was provided by the Muon Science Innovative Channel (MuSIC) - M1 beamline~\cite{Cook2017}. A graphite pion production target was irradiated with a primary proton beam accelerated to 392 MeV by two cyclotrons.
The primary beam intensity was 20~nA, which was only 2\% of a designed value of 1.1~$\mathrm{\mu A}$ because of radiation safety at the time of this experiment.
Negative pions produced at the graphite target were collected and transported by the pion capture solenoid and they decayed into negative muons.
These muons were momentum selected by the two dipole magnets of the beamline.
A Wien filter was used to remove electrons contaminating the beam. The beam contained 70\% muons and 30\% electrons after the Wien filter.
The muon momentum of 50~MeV/c was selected and decelerated to 40~MeV/c by a carbon degrader.
The beam had a 50~mm diameter in the full-width half-maximum, and the momentum spread was approximately 8.8\%~\cite{Tomono}.

Five isotopically enriched metal palladium targets were irradiated with the muon beam.
Each target is enriched to about 98\%, and a detailed composition is summarized in Table~\ref{tab:target}.
The $^{104}{\rm Pd}$, $^{108}{\rm Pd}$ and $^{110}{\rm Pd}$ target are metal discs with each diameter of about 15~mm and thickness of about 0.5~mm.
The $^{105}{\rm Pd}$ and $^{106}{\rm Pd}$ target are metal powder and encapsulated in graphite cases with a thickness of 1~mm for each side.
The inside dimensions of the graphite cases were 20~mm in diameter and 2.2~mm thickness for the $^{105}{\rm Pd}$ target, and 15~mm in diameter and 2.3~mm thickness for the $^{106}{\rm Pd}$ target, respectively~\cite{Terada2015}.

\begin{figure}[]
  \centering
  \includegraphics[width=8.0cm]{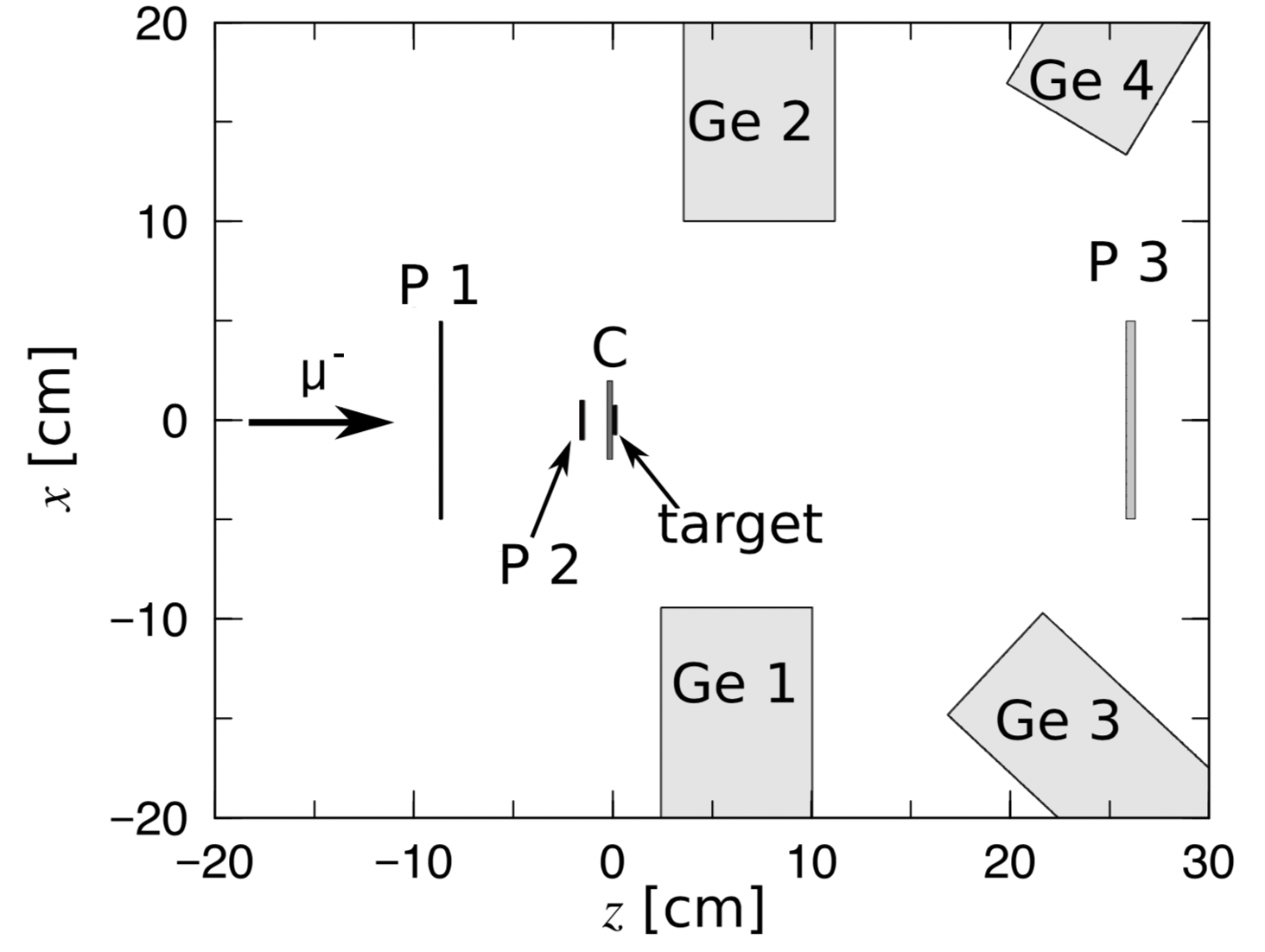}
  \caption{A top view of the detector configuration. The detectors are shown on the actual position and size, but the thickness of the upstream counters (P1 and P2) and the target are drawn thicker to see easily.}
  \label{fig:detector}
\end{figure}

A schematic drawing of the detector setup is shown in Fig.~\ref{fig:detector}.
The muon beam direction is along the $z$-axis.
The palladium target was attached on the downstream side of the carbon degrader~(C).
The thickness of the degrader was 3~mm for the disk targets and 2~mm for the powder targets considering the thickness of the graphite case.
Two beam counters~(P1, P2) were installed upstream of the palladium target, and one beam veto counter~(P3) was placed downstream.
The beam and veto counters consist of square-shaped plastic scintillators with photomultiplier-tube readout.
The effective area was $100 \times 100 $~${\rm mm^{2}}$, $20 \times 20$~${\rm mm^{2} }$ and $100 \times 100$~${\rm mm^{2} }$, and the thickness was 0.5~mm, 0.5~mm, and 5~mm for P1, P2, and P3, respectively.
X rays and $\gamma$ rays emitted from the palladium target were measured by four high-purity germanium detectors surrounding the target~(Ge1-4).
Ge1 and Ge2 were 30\% p-type coaxial detectors (CANBERRA GC3018) for high-energy photons, while Ge3 and Ge4 were n-type coaxial and planner type detectors (ORTEC GMX25195 and GLP36360), respectively, for low-energy photons.

The pulse heights of each germanium detector were recorded by a 13-bit peak-hold Analog-to-Digital Converter~(ADC) NIKI GLASS A3400.
The integrated charge of each beam counter was taken by a charge-sensitive ADC~(QDC) CAEN V792.
A Time-to-Digital Converter~(TDC) CAEN V1290 was used to measure the timing of the germanium detectors and the beam counters.
A data acquisition system based on RIBFDAQ~\citep{Baba2010} was used for data handling.
The data acquisition trigger was generated by the coincidence of P1 and P2.
The trigger rate was typically 30 particles per second.
The data was accumulated in 7.2~hours for the ${\rm ^{108}Pd}$, 3~hours each for the other palladium targets, and 1~hour for an empty measurement without targets, respectively. 
The empty measurement was used to estimate the $\gamma$-ray background.

\begin{figure}[]
  \centering
  \includegraphics[width=8.0cm]{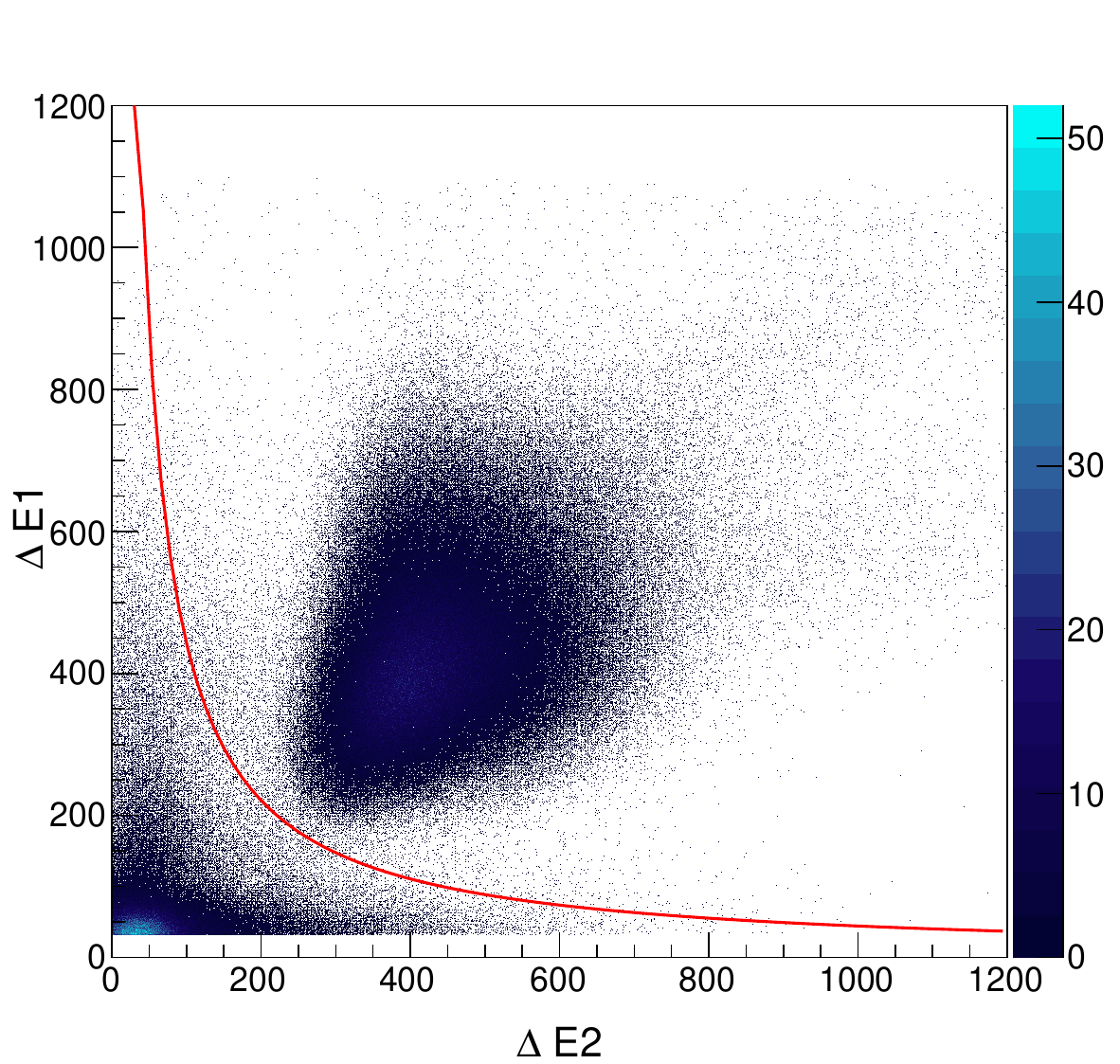}
  \caption{Particle identification with P1 and P2. The vertical and horizontal axes ($\Delta E_1$ and $\Delta E_2$) are the energy deposits measured by P1 and P2, respectively. The solid line represents the threshold value to discriminate the muon events from the electron events.}
  \label{fig:plaPI}
\end{figure}

The muons and electrons in the beam were identified by the energy loss at P1 and P2 because the muon deposits higher energy than the electron with the same momentum.
The energy deposits measured by P1 and P2, $\Delta E_1$ and $\Delta E_2$, respectively, are shown in Fig.~\ref{fig:plaPI}.
We adopted a threshold on the geometric mean of the QDC data of P1 and P2 to identify the muon as shown as the solid line in the figure.  
Some muons did not stop at the target and reached P3 since the effective area of P2 is larger than the size of the targets.
The events containing any signals at P3 were eliminated from the following analysis.

\begin{figure}[]
  \centering
  \includegraphics[width=7cm, angle = 0]{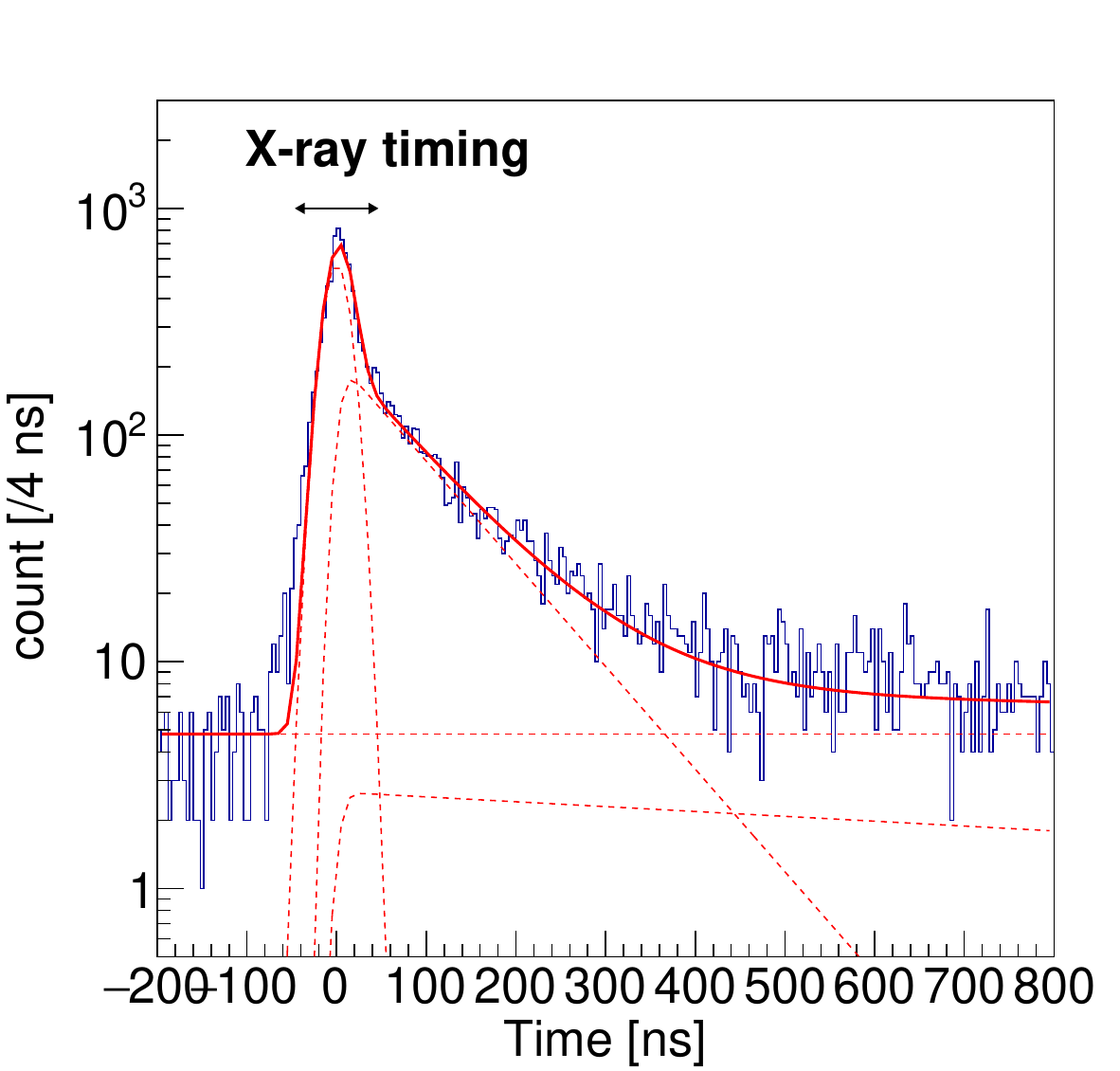}
  \caption{The timing spectrum of the photons after the muon irradiation to $\mathrm{^{108}Pd}$. The timing gate used for the X-ray spectrum is also shown. The red solid line represents the fitting function, Eq.~(\ref{eq:timefit}), and the dotted lines are each component in the function. See text for detail.}
  \label{fig:timing}
\end{figure}

After the muon stops at the target, the muonic atom is formed, and muonic X rays are emitted immediately.
The muon in the 1s orbit decays via $\mu$-$e$ decay or the muon capture with a lifetime of about 100~ns for the palladium.
The highly excited nucleus is produced via the muon capture, and it emits nucleons and $\gamma$ rays.
The X rays and the $\gamma$ rays can be separated by the photon detection timing.
Figure~\ref{fig:timing} shows the photon detection time spectrum at the germanium detectors against the beam detection timing at P2 for $\mathrm{^{108}Pd}$.
The prompt peak in the spectrum corresponds to the X ray, and the following exponential decay is originated from the $\gamma$ ray accompanied by the muon capture reaction.
To determine the lifetime of the muonic palladium, fitting of the time spectrum is performed.
The red solid curve in the figure is the following function ($f(t)$) used to fit the spectrum:
\begin{equation}
    f(t) = f_{X}(t) + f_\mathrm{Pd}(t) + f_\mathrm{C}(t) +  B, \label{eq:timefit}
\end{equation}
where $f_{X}$ and $f_\mathrm{Pd}$ are time contributions of the prompt X rays and the muon capture of Pd, respectively.
There are two background terms: $B$ is a constant background and $f_\mathrm{C}$ is the contribution of $\gamma$ rays and electrons produced by the muons that stopped at the carbon surrounding the target.
Each component is defined as 
\begin{align}   
    f_X(t) &= A_{X}\exp \left( - \frac{t^2}{2 \sigma_{T_{X}}^2} \right), \label{eq:timefit_x}\\
    f_\mathrm{Pd}(t) &=  A_\mathrm{Pd}\exp \left( - \frac{t}{\tau} \right)\left[1- \mathrm{Erf} \left(\frac{ (\sigma_{T_{\gamma}}^2/\tau) - t}{ \sqrt{2} \sigma_{T_{\gamma}} } \right) \right] \label{eq:timefit_gamma}\\
    f_\mathrm{C}(t) &=  A_\mathrm{C}\exp \left( - \frac{t}{\tau_C} \right)\left[1- \mathrm{Erf} \left(\frac{ (\sigma_{T_{\gamma}}^2/\tau_\mathrm{C}) - t}{ \sqrt{2} \sigma_{T_{\gamma}} } \right) \right] \label{eq:timefit_carbon}.
\end{align}
where $A_X$, $A_\mathrm{Pd}$, and $A_\mathrm{C}$ are the amplitudes of each component, $\mathrm{Erf}(x)$ is an error function, $\tau$ is the lifetime of the muonic palladium, $\tau_\mathrm{C}$ is that of the muonic carbon~\cite{Suzuki1987}, and $\sigma_{T_X}$ and $\sigma_{T_{\gamma}}$ represent the timing resolutions of the germanium detectors averaged in the energy region of the X rays and $\gamma$ rays, respectively.
The X-ray term is symmetric and two $\gamma$-ray terms are asymmetric considering their decay time convoluted with the detector timing resolution.
The error function is employed to express this asymmetry.
The fitting parameters are $\tau$, $A_X$, $A_\mathrm{Pd}$, $A_\mathrm{C}$, and $B$.
The obtained lifetimes of the palladium isotopes are summarized in Table~\ref{tab:lambdac}.
To obtain X-ray energy spectra, the photon timing is gated on the $\pm 3 \sigma_{T_X} $ prompt region.

For the energy calibration of the germanium detectors, standard $\gamma$-ray sources of $^{60}\mathrm{Co}$, $^{133}\mathrm{Ba}$, $^{137}\mathrm{Cs}$, and $^{152}\mathrm{Eu}$ were used. They emit the $\gamma$ rays with the energy from 80~keV to 1.3~MeV.
In addition to these sources, background $\gamma$ peaks at 2223~keV from $p(n,\gamma)d$ reaction, 3539~keV from $^{28}\mathrm{Si}(n,\gamma)^{29}\mathrm{Si}$ reaction, 2754~keV from $^{27}\mathrm{Al}(n,\alpha)^{24}\mathrm{Na}$ reaction and 2615~keV of $^{208}\mathrm{Tl}$ decay were used for the energy calibration in the high energy region.

\begin{figure*}[]
  \centering
  \includegraphics[width=16cm, angle = 0]{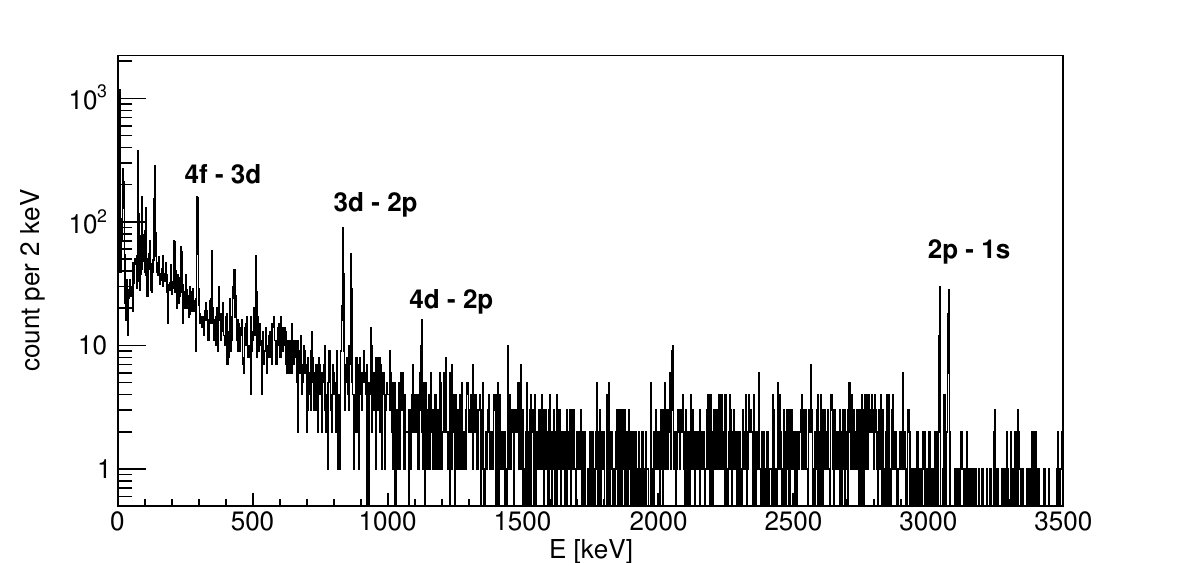}
  \caption{The entire spectrum of the $\mathrm{ ^{108}Pd}$ measurement gated on the X-ray timing. 
  The intense peaks are indicated.
  }
  \label{fig:xrayspectrum}
\end{figure*}

We report muonic X-ray energies up to $4f$-$3d$ transitions observed for five stable palladium isotopes for the first time.
Figure~\ref{fig:xrayspectrum} shows the X-ray energy spectrum for $^{108}\mathrm{Pd}$.
The peaks were assigned to each muonic X-ray transition and fitted by a Gaussian function with a constant background term.
The X-ray energies ($E$) of each transition are summarized in Table~\ref{tab:xray} for even isotopes and Table~\ref{tab:105Pd} for $^{105}$Pd.
For $^{105}\mathrm{Pd}$, the $2p_{3/2}$-$1s_{1/2}$ transition shows the hyperfine splitting, and it is impossible to estimate the detailed peak structure due to the lack of resolution and statistics.
The uncertainties in Table~\ref{tab:xray} and \ref{tab:105Pd} include statistical and systematic uncertainties.
The systematic uncertainty on the energy is caused by the non-linearity of the circuits (the pre-amplifier and ADC) and the gain drift of the germanium detectors during the experiment.
Note that the statistical uncertainty is dominant for the $2p$-$1s$ transitions, while the systematic uncertainty is the major component for the other transitions.
The $2p$-$1s$ transition energies appear in the compilations~\cite{Fricke1995,Angeli2013}, though the original paper was unpublished~\cite{Hack1989}, and the obtained energies are consistent with the values in the compilation within the uncertainty.

\begin{table}[]
  \centering
  \caption{The lifetimes ($\tau$) and the total capture rates ($\Lambda_C$) of muonic palladium determined in this work.
    Theoretical total capture rates are taken from Zinner \textit{et al.}~\citep{Zinner2006} and Marketin \textit{et al.}~\citep{Marketin2009}. See text for detail of the calculation.}
  \label{tab:lambdac}
  \begin{tabular}{lcccc} \hline \hline
    & $\tau$ [ns] & $\Lambda_C$~$[10^6~{\rm s^{-1}}]$ &  Zinner & Marketin \\\hline
    $^{104}{\rm Pd}$ & $81 (3)$  &$12.0 (4)$ & 12.71 & 13.182  \\
    $^{105}{\rm Pd}$ & $92 (6)$ &$10.5 (7)$  &  - & -\\
    $^{106}{\rm Pd}$ & $85 (4)$  &$11.4 (6)$  & 11.44 & 11.912  \\
    $^{108}{\rm Pd}$ & $96 (3)$ &$10.0 (3)$  & 10.44 & 10.746  \\
    $^{110}{\rm Pd}$ & $94 (6)$ &$10.2 (6)$  & 9.607 & 9.821  \\
    \hline \hline
  \end{tabular}
\end{table}

\begin{table*}[]
  \centering
  \caption{Observed muonic X-ray energies for even palladium isotopes. 
  The Fermi parameters $c$ and $t$ determined by the transition energies are also listed. 
  The Fermi parameters $c$ in the reference which is determined with fixed $t=2.3$~fm are also shown.
  The Barrett parameters $k$,$\alpha_B$, and $R_{k\alpha_B}$ are also listed.}
  \begin{tabular}{l @{\hspace{0.5cm}} l  cc @{\hspace{0.5cm}}  ccc ccc cc}
    \hline \hline
    Nuclei     & \multicolumn{2}{c}{X-ray energies} &   \multicolumn{6}{c}{Charge parameters from present experiment}  & \multicolumn{2}{c}{Reference~\cite{Fricke1995}} \\
               & transition               &\quad $E$ [keV] \quad \quad & $k$   & $\alpha_B$ [fm$^{-1}$] & $R_{k\alpha_B}$ [fm] & $c$ [fm] & $t$ [fm] & rms [fm] &\quad \quad $c$ [fm] \quad &  rms [fm]   \\
 \hline
    $^{104}$Pd &  $2p_{1/2} $-$ 1s_{1/2}$ & 3057.8(3) & 2.163 & 0.1037 & 5.762(2) & 5.26(12) & 2.25(25) & 4.493$_{-7}^{+28}$ & 5.2509(4) & 4.509 \\
               &  $2p_{3/2} $-$ 1s_{1/2}$ & 3091.7(2) & 2.145 & 0.1013 & 5.764(2) & & & & & \\
               &  $3d_{5/2} $-$ 2p_{3/2}$ & 833.4(1)  & 3.705 & 0.0802 & 5.91(2)  & & & & & \\
               &  $3d_{3/2} $-$ 2p_{1/2}$ & 863.6(3)  & 3.798 & 0.1183 & 5.89(4)  & & & & & \\
               &  $4d_{5/2} $-$ 2p_{3/2}$ & 1122.5(6) & 3.703 & 0.0795 & 5.9(1)   & & & & & \\
               &  $4f_{7/2} $-$ 3d_{5/2}$ & 291.6(1)  & 5.864 & 0.1257 & 6.1(4)   & & & & & \\
               &  $4f_{5/2} $-$ 3d_{3/2}$ & 294.9(2)  & 6.252 & 0.0963 & 6.3(6)   & & & & & \\

 \hline
    $^{106}$Pd &  $2p_{1/2} $-$ 1s_{1/2}$ & 3050.3(5) & 2.157 & 0.1027 & 5.792(2) & 5.49(30) & 1.76(65) & 4.507$_{-5}^{+96}$ & 5.2847(2) & 4.532 \\ 
               &  $2p_{3/2} $-$ 1s_{1/2}$ & 3084.0(3) & 2.148 & 0.1013 & 5.795(2) &     &     &    &&   \\
               &  $3d_{5/2} $-$ 2p_{3/2}$ & 833.4(3)  & 3.705 & 0.0802 & 5.92(7)  &     &     &    &&   \\
               &  $3d_{3/2} $-$ 2p_{1/2}$ & 863.8(3)  & 3.798 & 0.1184 & 5.86(4)  &     &     &    &&   \\
               &  $4d_{5/2} $-$ 2p_{3/2}$ & 1124.8(4) & 3.703 & 0.0795 & 5.77(9)  &     &     &    &&   \\
               &  $4f_{7/2} $-$ 3d_{5/2}$ & 291.7(2)  & 5.864 & 0.1257 & 5.6(7)   &     &     &    &&   \\
               &  $4f_{5/2} $-$ 3d_{3/2}$ & 295.1(3)  & 6.251 & 0.0963 & 5.6(8)   &     &     &    &&   \\
 
 \hline 
    $^{108}$Pd &  $2p_{1/2} $-$ 1s_{1/2}$ & 3042.9(2) & 2.153 & 0.1020 & 5.822(2) &  5.45(11)   & 1.96(27)  & 4.537$_{-6}^{+27}$ &  5.3184(2) & 4.556 \\  
               &  $2p_{3/2} $-$ 1s_{1/2}$ & 3076.6(2) & 2.149 & 0.1012 & 5.824(2) &     &     &    &&   \\  
               &  $3d_{5/2} $-$ 2p_{3/2}$ & 833.3(1)  & 3.705 & 0.0799 & 5.95(2)  &     &     &    &&   \\   
               &  $3d_{3/2} $-$ 2p_{1/2}$ & 863.5(1)  & 3.798 & 0.1184 & 5.91(1)  &     &     &    &&   \\  
               &  $4d_{5/2} $-$ 2p_{3/2}$ & 1124.4(3) & 3.702 & 0.0793 & 5.87(7)  &     &     &    &&   \\   
               &  $4f_{7/2} $-$ 3d_{5/2}$ & 291.6(1)  & 5.864 & 0.1257 & 5.8(3)   &     &     &    &&   \\   
               &  $4f_{5/2} $-$ 3d_{3/2}$ & 294.9(1)  & 6.252 & 0.0963 & 6.3(3)   &     &     &    &&   \\   

 \hline
    $^{110}$Pd &  $2p_{1/2} $-$ 1s_{1/2}$ & 3036.2(7) & 2.149 & 0.1013 & 5.849(3) & 5.60(26)    & 1.61(70)  & 4.548$_{-6}^{+100}$ & 5.3490(3) & 4.577\\ 
               &  $2p_{3/2} $-$ 1s_{1/2}$ & 3068.8(4) & 2.153 & 0.1013 & 5.856(2) &     &     &    &&   \\ 
               &  $3d_{5/2} $-$ 2p_{3/2}$ & 833.4(2)  & 3.704 & 0.0799 & 5.94(5)  &     &     &    &&   \\  
               &  $3d_{3/2} $-$ 2p_{1/2}$ & 863.4(3)  & 3.798 & 0.1184 & 5.92(4)  &     &     &    &&   \\ 
               &  $4d_{5/2} $-$ 2p_{3/2}$ & 1124.2(5) & 3.702 & 0.0793 & 5.9(1)   &     &     &    &&   \\  
               &  $4f_{7/2} $-$ 3d_{5/2}$ & 291.6(1)  & 5.864 & 0.1257 & 6.1(4)   &     &     &    &&   \\  
               &  $4f_{5/2} $-$ 3d_{3/2}$ & 294.9(2)  & 6.252 & 0.0963 & 6.3(8)   &     &     &    &&    \\  

 \hline \hline

  \end{tabular}
  \label{tab:xray}
\end{table*}

\begin{table}[]
  \centering
  \caption{Observed muonic X-ray transition energies for $^{105}$Pd. }
  \label{tab:105Pd}
  \begin{tabular}{lc} \hline \hline
    \multicolumn{2}{c}{X-ray energies}\\
    transition & E [keV] \\
    \hline
    $2p_{1/2} $-$ 1s_{1/2}$&3054.9(5)  \\
    $2p_{3/2} $-$ 1s_{1/2}$&3091.5(8)   \\
    $3d_{5/2} $-$ 2p_{3/2}$&832.3(3)  \\
    $3d_{3/2} $-$ 2p_{1/2}$&864.5(3)  \\
    $4f_{7/2} $-$ 3d_{5/2}$&291.4(2)  \\
    $4f_{5/2} $-$ 3d_{3/2}$&294.9(6)  \\

   \hline \hline
  \end{tabular}
\end{table}

\section{Total muon capture rate}\label{sec_lambdac}

The total capture rate of the nuclear muon capture ($\Lambda_C$) corresponds to the lifetime of the muonic $1s$ state ($\tau$) with the following relation:
\begin{equation}
\frac{1}{\tau} = \Lambda_C + Q \frac{1}{\tau_{\mu^+}},
\end{equation}
where $\tau_{\mu^+}$ is the lifetime of a positive muon (2.1969811(22)~$\mu$s)~\cite{PDG2014} and $Q$ is the Huff factor. The Huff factor is $Q=0.927$ for palladium~\citep{Suzuki1987,Blair1962}. 

The total muon capture rates deduced from the measured lifetimes are listed in Table~\ref{tab:lambdac} for each palladium isotope.
In the fitting procedure, we assume the lifetimes of excited states in each rhodium isotope produced by the muon capture are negligibly small. If the excited states have long lifetimes, the measured $\tau$ becomes longer than the actual lifetime of the muonic palladium. Hence, the values of $\tau$ in Table~\ref{tab:lambdac} are the upper limits. Consequently, the capture rates $\Lambda_C$ shown here are the lower limits.

For a consistency check, the lifetime of natural palladium was calculated by summing up the measured decay curves with the weight of the natural abundance of palladium. 
Natural palladium contains 1.02\% of $\mathrm{^{102}Pd}$ and the existence of $\mathrm{^{102}Pd}$ is neglected here.
The lifetime of this accumulated decay curve is $90 \pm 5$~ns, which is consistent with the previously measured value $96 \pm 0.6$~ns~\citep{Eckhause1966} of the natural palladium within the uncertainty.

The nuclear muon capture changes a proton into a neutron in the excited states.
The excited states following the nuclear muon capture reaction can be well described by the random phase approximation (RPA), in which each excited state is written as a superposition of one-particle-one-hole states. The theoretical calculation of $\Lambda_C$ is also listed in Table~\ref{tab:lambdac} with RPA based on single-particle states generated by Woods-Saxon potential by Zinner \textit{et al.}~\citep{Zinner2006} and the relativistic proton-neutron quasiparticle random phase approximation ({\it pn}-RQRPA) using the relativistic Hartree-Bogoliubov basis by Marketin \textit{et al.}~\citep{Marketin2009}. 
The RPA-based calculations well reproduced the present experimental data.


\section{Discussion I: Nuclear Charge Radius}\label{sec_radius}
In this section, the nuclear charge radii of the palladium isotopes are deduced from the measured muonic transition energies.
The two-parameter Fermi (2pF) function is assumed for the charge distribution in the analysis.
Once the 2pF function is assumed, the two parameters in the 2pF function are determined using the experimental inputs, and then the rms charge radius can be calculated from the 2pF function.

Here we follow references~\cite{Barrett1974, Borie1982, Michel2017}.
One can start from a simple two-body bound system with the muon and the nucleus in their center-of-mass system using their reduced mass ($m_{r}$).
Omitting the nuclear excitation, the Hamiltonian of the muonic atom can be written as
\begin{equation}
  H = H_{\mu} + H_{\mu-N},
  \label{eq:hamiltonian}
\end{equation}
where $H_{\mu}$ is the free Dirac Hamiltonian with the reduced mass ($m_r$) and $H_{\mu-N}$ is the interaction Hamiltonian.
Considering the ground state of a spherical even-even nucleus with a spin parity of $0^+$, $H_{\mu-N}$ contains only the static electric part.
The interaction Hamiltonian can be expressed with the fine structure constant $\alpha$ and the nuclear charge distribution $\rho(r)$ as
\begin{equation}
  H_{\mu-N} = -\alpha \int \frac{\rho(r')}{|r-r'|} \mathrm{d}r',
  \label{eq:static_potential}
\end{equation}
where $\rho(r)$ must be normalized to be $\int \rho \mathrm{d}r^3 = Z$.
For $0^+$ state, the static potential with the spherical charge distribution (Eq.~(\ref{eq:static_potential})) is also adequate for deformed nuclei as the first approximation. 
This approximation is also justified for deformed nuclei considering a time-averaged distribution.

For the numerical calculation, one must assume the functional form of $\rho(r)$.
The 2pF distribution with parameters $c$ and $t$ is used in this analysis:
\begin{equation}
  \rho_{\mathrm{2pF}}(r) = \frac{ N_0 }{1 + \exp{[ \frac{(r-c)}{t/(4 \ln{3}) } ]}},
\label{eq:2pF}
\end{equation}
where $N_0$ is the normalization factor.
The Fermi parameters, $c$ and $t$, are the half-density radius and the diffuseness, respectively.
Once the functional form of $\rho(r)$ is given, the binding energies of the muonic atom are calculated by numerically solving the eigenvalue problem of the Hamiltonian (Eq.~(\ref{eq:hamiltonian})).
The muonic X-ray transition energies are then calculated to be the difference of the binding energies between the two corresponding states.

For further correction, the QED and relativistic recoil corrections should be included in the bare wavefunctions and binding energy of Eq.~(\ref{eq:hamiltonian}) as same as an ordinal electronic atom.
For the muonic atom, the electron screening effect by the inner shell electrons should also be taken into account.
We include these perturbative corrections following Ref.~\cite{Michel2017}.

In addition to the above corrections, an energy shift by nuclear polarization and self-energy correction should also be considered.
The binding energy of the muonic atom reaches several MeV, which is comparable with the nuclear excitation energy.
The coupling between the nuclear structure and the muonic atomic states causes an energy shift by several keV in the binding energy.
This energy shift is called a nuclear polarization effect.
Complete calculation of the nuclear polarization correction is almost impossible because all excited states in the nucleus contribute to the energy shift.
The energy shift from the nuclear polarization is estimated to be about 1.4~keV for $2p$-$1s$ transitions for the palladium isotopes~\cite{Fricke1995}.
The energy shift from the self-energy correction is relatively smaller than the electronic atom and estimated to be the order of 1~keV for the $1s$ state~\cite{Cheng1978}.
The nuclear polarization effect and self-energy correction are included by interpolation of the previous data~\cite{Fricke1995}. 
The uncertainty due to the theoretical calculation is not included in the following analysis.



\begin{figure}[]
  \centering
 \includegraphics[width=7cm, angle = 0]{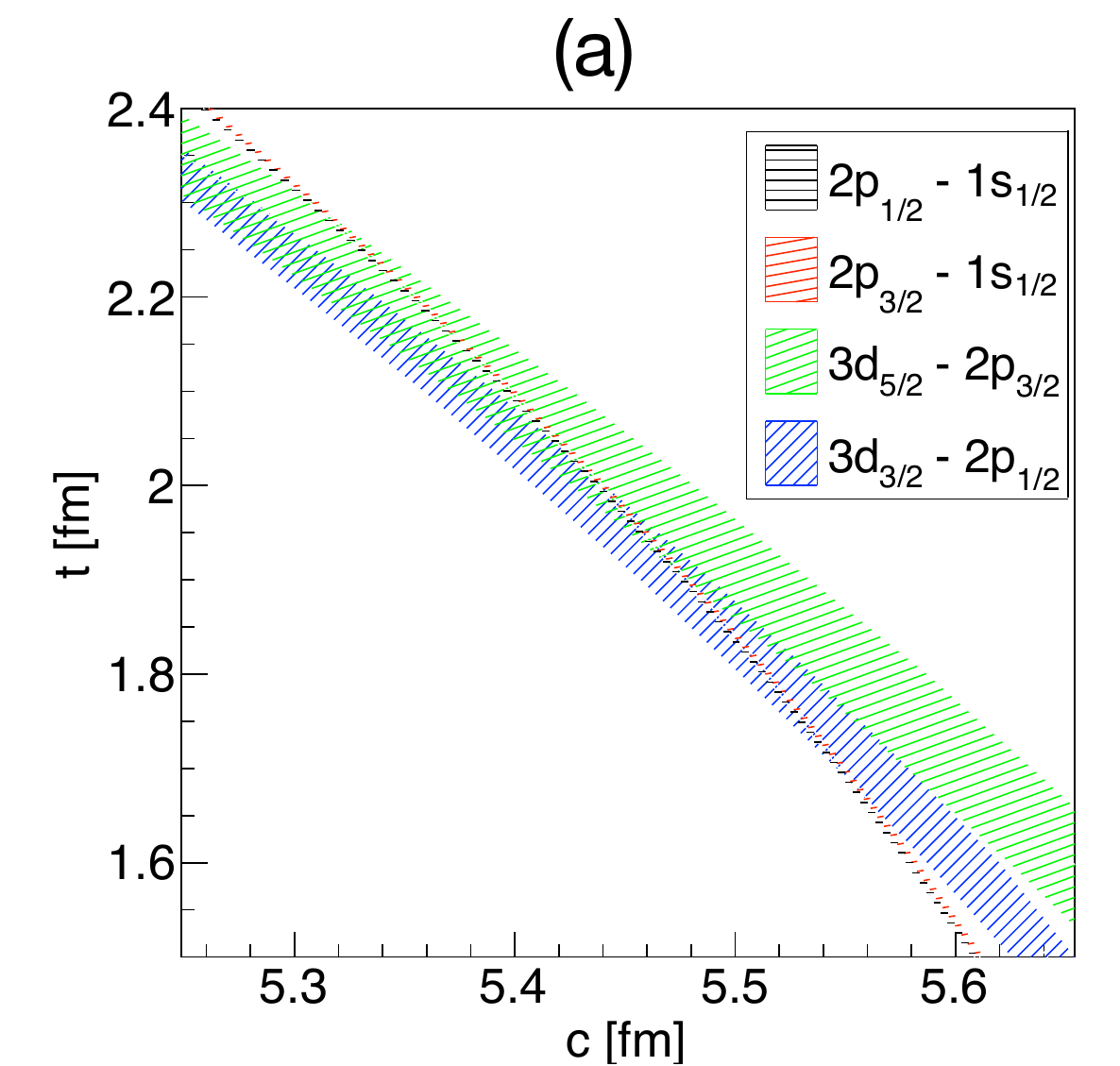}
 \includegraphics[width=7cm, angle = 0]{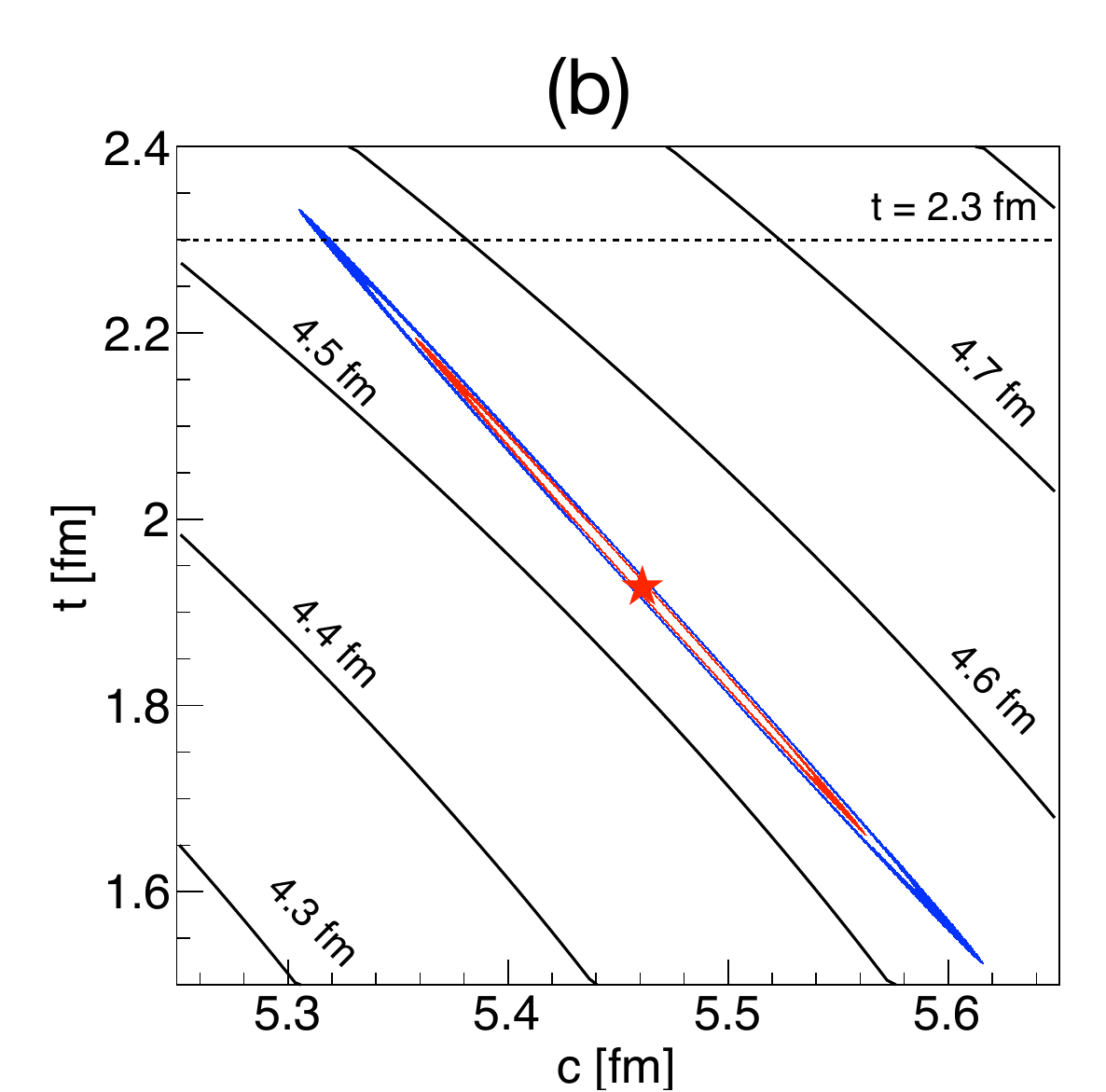}
  \caption{ (a) 1$\sigma$ region limited by the obtained muonic transition energies in the two-dimensional $c$-$t$ space for $^{108}$Pd. Each shaded area corresponds to limited region by each muonic transition with 1$\sigma$ uncertainty. (b) 1$\sigma$ uncertainty region of the Fermi parameter in the $c$-$t$ space for $^{108}{\rm Pd}$. The red star symbol shows the determined Fermi parameters which gives the $\chi^2$ minimum. The red and blue ellipses represent the region for $\chi^2 = \chi^2_{\mathrm{min}}+ 1$ and $\chi^2 = \chi^2_{\mathrm{min}}+ 2.3$, respectively. The contours of the rms radii (0.1~fm interval) are illustrated with the black solid lines. The black dotted line represents $t = 2.3$~fm.
  }
    \label{fig:fermi2d}     
\end{figure}

The observed X-ray energies in Table~\ref{tab:xray} are different from the transition energies because they are affected by the photon recoil effect, in which the atom is kicked by the X ray and a part of the energy of the X ray is transferred to the atom. 
The experimental transition energies were obtained from the observed X-ray energies by applying the photon recoil correction.
The photon recoil correction is calculated to be 0.04~keV for $2p$-$1s$ transitions at 3~MeV and negligibly small for the other transitions (about 0.005~keV for $3d$-$2p$ transition for example). 

To obtain the Fermi parameters, the theoretically calculated transition energies were fit to the experimental transition energy through the $\chi^2$ minimization procedure.
It would be important to note that the single transition energy is only able to limit the two-dimensional $c$-$t$ parameter space ($c$-$t$ space) as a belt-like area corresponding to its experimental uncertainty, as shown in Fig.~\ref{fig:fermi2d}-(a).
The combination of more than two areas with different gradients in the $c$-$t$ space is required to determine the pair of the Fermi parameters. 
It is thus necessary to include the higher transitions adding to the lowest transitions $2p_{1/2}$-$1s_{1/2}$ and $2p_{3/2}$-$1s_{1/2}$ since these two transitions are completely overlapped to be distinguished.
Therefore, we include four observed transitions, $2p_{1/2}$-$1s_{1/2}$, $2p_{3/2}$-$1s_{1/2}$, $3d_{3/2}$-$2p_{1/2}$ and $3d_{5/2}$-$2p_{3/2}$, to the minimization procedure.
The other transitions were not used because of their low statistics.
The parameters $c$ and $t$ were scanned independently to find the minimum value of $\chi^2$ ($\chi^2_\mathrm{min}$) in the $c$-$t$ space.
The determined Fermi parameters are summarized in Table~\ref{tab:xray} together with their uncertainty.

The rms radii were calculated as the square root of the second-order nuclear moment from the obtained parameters;
\begin{equation}
    \mathrm{(rms)}^2 = \langle r^2\rangle =  \int \rho_{\mathrm{2pF}} (r) 4 \pi r ^2 \mathrm{d}r.
\end{equation}
The deduced rms radii for the palladium isotopes are listed in Table~\ref{tab:xray}.
The 1$\sigma$ uncertainty region, namely the region in which the $\chi^2$  is less than $\chi^2_\mathrm{min} + 1$ for one parameter and $\chi^2_\mathrm{min} + 2.3$ for two parameters, in the $c$-$t$ space is illustrated for $^{108}$Pd in Fig.~\ref{fig:fermi2d}-(b).
The uncertainties quoted for the Fermi parameters are taken from $\chi^2_{\mathrm{min}} + 1$ region and the uncertainties for rms radii are estimated from $\chi^2_{\mathrm{min}} + 2.3$ region. 
The uncertainty in the extracted rms radius does not account for the theoretical uncertainty associated with muonic atom energy levels. 
Instead, the uncertainty presented here reflects the potential precision of our method in determining charge radii from measured transition energies. 
The upper and lower uncertainties are estimated by calculating the maximum and minimum rms radius within the 1-sigma region shown in Fig.~\ref{fig:fermi2d}-(b).
The asymmetry between the upper and lower uncertainties can be explained by the projection procedure from the two-dimensional c-t space to the one-dimensional rms radius.

The uncertainty quoted for the rms radii in Table~\ref{tab:xray} is relatively small compared to that for the Fermi parameters.
This is because the sensitivity of the X-ray energies to the rms radius is higher than that of the Fermi parameters.
The high sensitivity to the rms radius is indicated by the fact that the experimental 1$\sigma$ region has a similar gradient to the contour of the rms radius as shown in Fig.~\ref{fig:fermi2d}-(b).
The transition energy of the $2p$-$1s$ transitions is sensitive to the rms radius, not to the Fermi parameters $c$ and $t$.



The Fermi parameters and the rms radii listed in the compilation are also listed on Table~\ref{tab:xray}.
The Fermi parameter $c$ in the reference~\cite{Fricke1995} seems much higher precision than the present results. 
This discrepancy is due to the different approaches used in the interpretation. 
In our interpretation, two Fermi parameters ($c$ and $t$) are treated as free parameters and determined by the measured X-ray energies including higher transitions. 
In the compilation, $c$ is deduced from only the $2p$-$1s$ transitions with the assumption of the fixed $t$ value as 2.3~fm. 
This fixed $t$ is an empirical value based on the systematics.
This assumption corresponds to the projection to the $t$ one-dimensional space at $t=2.3$ in Fig.~\ref{fig:fermi2d}-(b) and leads an extremely small uncertainty. 
It should be noted that no uncertainty for the rms radii is given in the compilation because of the difficulty to estimate the systematic uncertainty from the fixed $t$ value.
Nevertheless, the rms radii obtained by the present work and the compilation are consistent with our uncertainty.

%

We can now discuss the accuracy of the rms radii.
In the present analysis, the uncertainty of the rms radius is limited by the experimental uncertainty of the $3d$-$2p$ transition energies.
The accuracy of the rms radius from each transition corresponds to the width of the belt-like area in Fig.~\ref{fig:fermi2d}-(a).
The accuracy of the rms radius from the $2p$-$1s$ transitions is higher about ten times than that from the $3d$-$2p$ transitions. 
The experimental uncertainty of the transition energies differs only twice for these transitions, namely 0.0066\% ($0.2$~keV for $3043$-keV X ray) and 0.012\% ($0.1$~keV for $833$-keV X ray) for the $2p$-$1s$ and $3d$-$2p$ transitions, respectively.
The difference in the accuracy thus reflects the difference in the intrinsic sensitivity for the nuclear charge radius: the change of the nuclear charge distribution affects the $2p$-$1s$ transition energy more drastically than the $3d$-$2p$ transitions since the former occurs closer to the nucleus than the latter.
In general, the higher transitions have less intrinsic sensitivity for the nuclear charge radius.
In order to compensate for the less intrinsic sensitivity, the energies of the $3d$-$2p$ transitions must be determined more precisely than those of the $2p$-$1s$ transitions for the accurate determination of the rms radius.
It should be mentioned again that the experimental uncertainties of the $3d$-$2p$ transition energies are dominated by the systematic uncertainty, which is mainly due to the non-linearity of the circuit, not by the statistical uncertainty.
For a future experiment, it is thus important to employ a sophisticated detector system to reduce the systematic uncertainty caused by the gain drift of the detector and the non-linearity of the circuit.
The uncertainty in the extracted rms radius does not account for the theoretical uncertainty associated with muonic atom energy levels. Instead, the uncertainty presented here reflects the potential precision of our method in determining charge radii from measured transition energies. The upper and lower uncertainties are estimated by calculating the maximum and minimum rms radius within the 1$\sigma$ region shown in Fig.~\ref{fig:fermi2d}-(b). The asymmetry between the upper and lower uncertainties can be explained by the projection procedure from the two-dimensional $c$-$t$ space to the one-dimensional rms radius.
The reader can find related discussions about the model uncertainty from a theoretical point of view in this paper~\cite{Xie2023}.

\section{Discussion II: Nuclear Charge Distribution}\label{sec_barret}
We discuss the nuclear charge distribution with the muonic transition energies in this section.
The major concern of the approach with the 2pF distribution is the limitation of the functional form.
Generally, the actual nuclear charge distribution $\rho(r)$ is complicated and has a non-analytic form, and any approach assuming a certain functional form must contain the limitation.
Thus, we employ the Barrett model, which is sometimes referred to as the ''model-independent'' approach and is motivated to treat the nuclear distribution without any functional assumptions~\cite{Barrett1970}.
In the Barrett model, a certain nuclear moment, called a Barrett moment, is derived as the perturbation from the distribution with an analytic functional form.
For the derivation of the Barrett moment, we follow Refs.~\cite{Barrett1970, Ford1973, Angeli2002, Engfer1974, Fricke1995}.


Let us begin with a reference nuclear charge distribution $\rho_0 (r)$ that is close to $\rho(r)$.
$\rho_0 (r)$ should be an analytic form to calculate the binding energy and wavefunctions and fulfill the following normalization; $\int \rho (r)\mathrm{d}^3 r =\int \rho_0(r) \mathrm{d}^3 r = Z $.
The muonic transition energy $E_{ij}[\rho]$ is the energy difference from the initial state $i$ to the final state $j$ and depends on the charge distribution $\rho (r)$.
In the first-order perturbation, the difference of the transition energies caused by the change of the charge distribution $\delta \rho(r) = \rho(r) - \rho_0(r)$ can be approximated as
\begin{align}
    \delta E &= E_{ij} [\rho] - E_{ij} [\rho_0] \notag \\
    &=\int \delta \rho(r) \Delta V_{ij}(r) 4\pi r^2 \mathrm{d}r. \label{eq:deltaEapprox}
\end{align}
$\Delta V_{ij}(r) = V_{i}(r) - V_{j}(r)$, where $V_{i}(r)$ is the Coulomb potential of the muon at the state $i$. It can be written as 
\begin{equation}
  \Delta V_{ij}(r) = \alpha \int \frac{ \psi_i^{\dagger}(r') \psi_i(r') - \psi_j^{\dagger}(r') \psi_j(r') }{ |r' - r| } 4 \pi r^2 \mathrm{d} r  
\label{eq:barrettpotential}
\end{equation}
using the muon wavefunctions $\psi_i(r)$ in each atomic state.
This treatment has also been applied for the analysis of isotope or isomeric shifts as considering $\delta \rho (r)$ as the distribution difference between the nuclei.
In this paper, $\delta \rho (r)$ always represents the difference between the reference distribution $\rho_0 (r)$ and the actual distribution $\rho (r)$.

For further simplification, $\Delta V_{ij}(r)$ is approximated with an analytic function.
Barrett proposed a phenomenological functional form 
\begin{equation}
  \Delta V_{ij} (r) \simeq A + B r^k e^{- \alpha_B r}
  \label{eq:barrettfunction}
\end{equation}
with four parameters $k$, $\alpha_B$, $A$, and $B$. 
The $\alpha_B$ is denoted with $\alpha$ in the references~\cite{Barrett1970, Ford1973, Angeli2002, Engfer1974, Fricke1995} and we use $\alpha_B$ in this paper to distinguish it from the fine-structure constant $\alpha$.
The form Eq.~(\ref{eq:barrettfunction}) is only valid in the limited region with $r < 10$-$20$~fm for typical cases and the product $\delta \rho (r) \Delta V_{ij}(r)$ becomes zero for large $r$. 
One can assume $\rho (r)$ and $\rho_0 (r)$ give the same $k$, $\alpha_B$, $A$, and $B$ as long as $\delta \rho (r)$ is small. 
Then these parameters are independent of the choice of the reference distribution $\rho_0 (r)$.
We can obtain the parameters by fitting Eq.~(\ref{eq:barrettfunction}) to the $\Delta V_{ij} (r)$ calculated using Eq.~(\ref{eq:barrettpotential}) with assumed $\rho_0 (r)$.
Note that the parameters are different among transitions even in the same nucleus.
Since $\delta \rho (r)$ is normalized to $\int \delta \rho (r) 4\pi r^2 \mathrm{d}^3 r= 0$, the constant term $A$ has no effect in the discussion below.

Using the parameters in (\ref{eq:barrettfunction}), the Barrett moment $ \langle r^k e^{- \alpha_B r} \rangle $ can be defined as
\begin{equation}
   \langle r^k e^{- \alpha_B r} \rangle  = \frac{1}{Z} \int \rho (r) r^k e^{- \alpha_B r} 4 \pi r ^2 \mathrm{d} r.
\label{eq:barrettmoment}
\end{equation}
Now the energy difference Eq.~(\ref{eq:deltaEapprox}) can be expressed as
\begin{equation}
  \delta E = Z B [ \langle r^k e^{- \alpha_B r} \rangle  - \langle r^k e^{- \alpha_B r} \rangle_0 ],
  \label{eq:barretttransition}
\end{equation}
where $\langle r^k e^{- \alpha_B r} \rangle_0$ denotes the Barrett moment with $\rho_0$.
Furthermore, the Barrett equivalent radius $R_{k\alpha_B}$ is introduced as follows:
\begin{equation}
  \frac{3}{R_{k\alpha_B}^3} \int_0 ^{R_{k\alpha_B}} r^k e^{- \alpha_B r} r ^2 \mathrm{d} r = \langle r^k e^{- \alpha_B r}\rangle.
\end{equation}
In this approach, derivation of the rms radius is difficult, and the equivalent radius is generally used for the comparison among the experimental results from the measurements of the muonic X rays.
The equivalent radius is independent of the parameter $B$ and depends only on $k$ and $\alpha_B$.
To determine the experimental $R_{k \alpha_B}$, the first-order form
\begin{equation}
  \delta R_{k \alpha_B } = C_Z \delta E
  \label{eq:firstorderCz}
\end{equation}
can be used.
The correction factor $C_Z$ is given by
\begin{equation}
   C_Z  = \frac{R_{k\alpha_B} }{ 3ZB [ \langle r^k e^{- \alpha_B r} \rangle_0 - R_{k \alpha_B}^k e^{- \alpha_B R_{k \alpha_B}} ] }
\end{equation}
with the equivalent radius $R_{k\alpha_B}$ that is calculated based on $\rho_0(r)$.

In Table~\ref{tab:xray}, $k$, $\alpha_B$, and $R_{k\alpha_B}$ calculated using $\rho_0 (r) = \rho_\mathrm{2pF} (r)$ with the Fermi parameters obtained in Sec.~\ref{sec_radius} are shown.
The QED, recoil, and electron screening corrections were included in this calculation.
For the uncertainty of $R_{k \alpha_B}$, the first-order form Eq.~(\ref{eq:firstorderCz}) is combined with the experimental uncertainty except for the $4f$-$3d$ transitions.
Because the experimental uncertainty is too large for the $4f$-$3d$ transitions to justify the first-order approximation, the uncertainty for these transitions is estimated by scanning the Fermi parameters within the experimental uncertainty and taking a conservative value.

\begin{figure}[t]
  \centering
 \includegraphics[width=7cm, angle = 0]{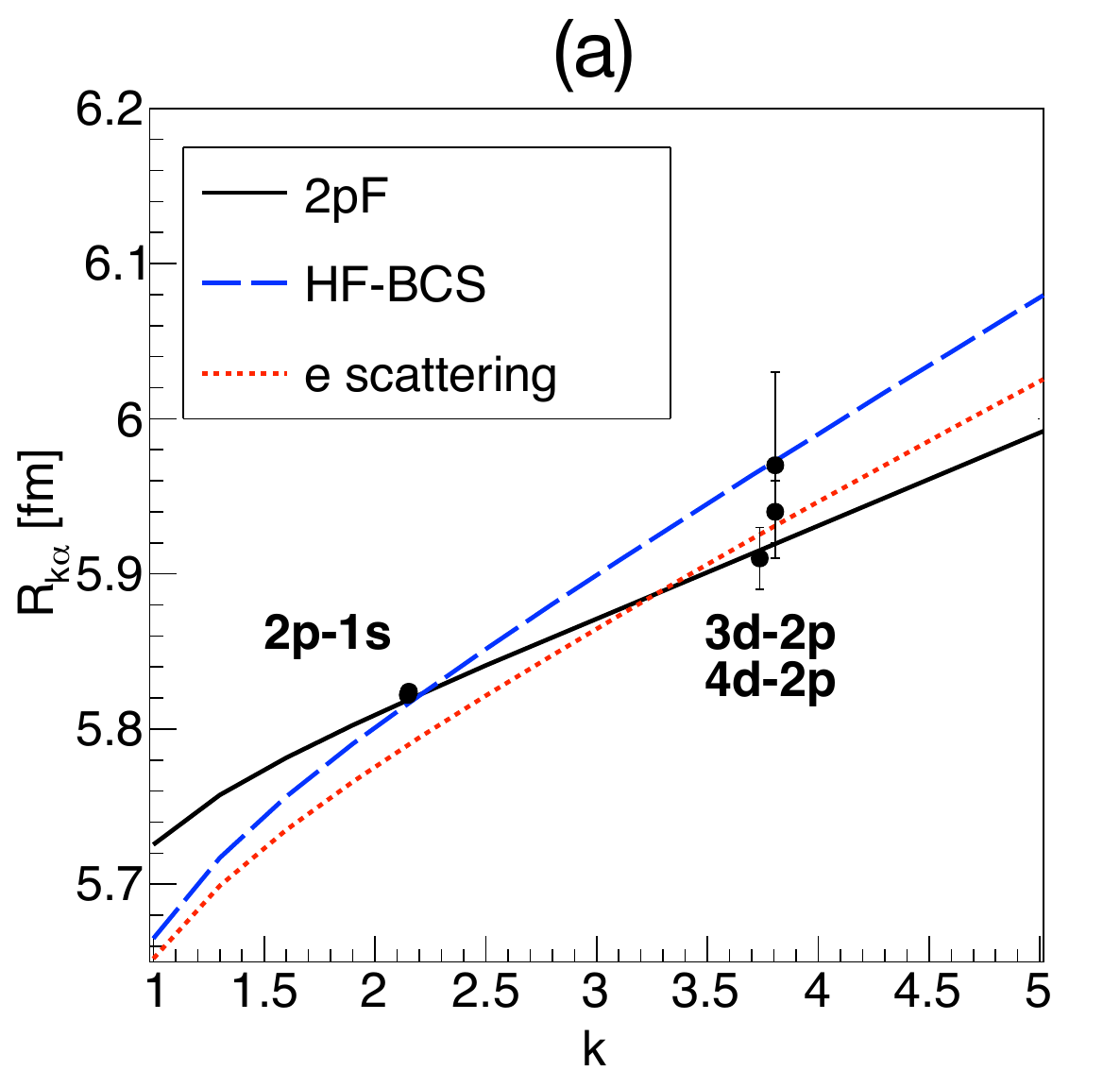} \\
 \includegraphics[width=7cm, angle = 0]{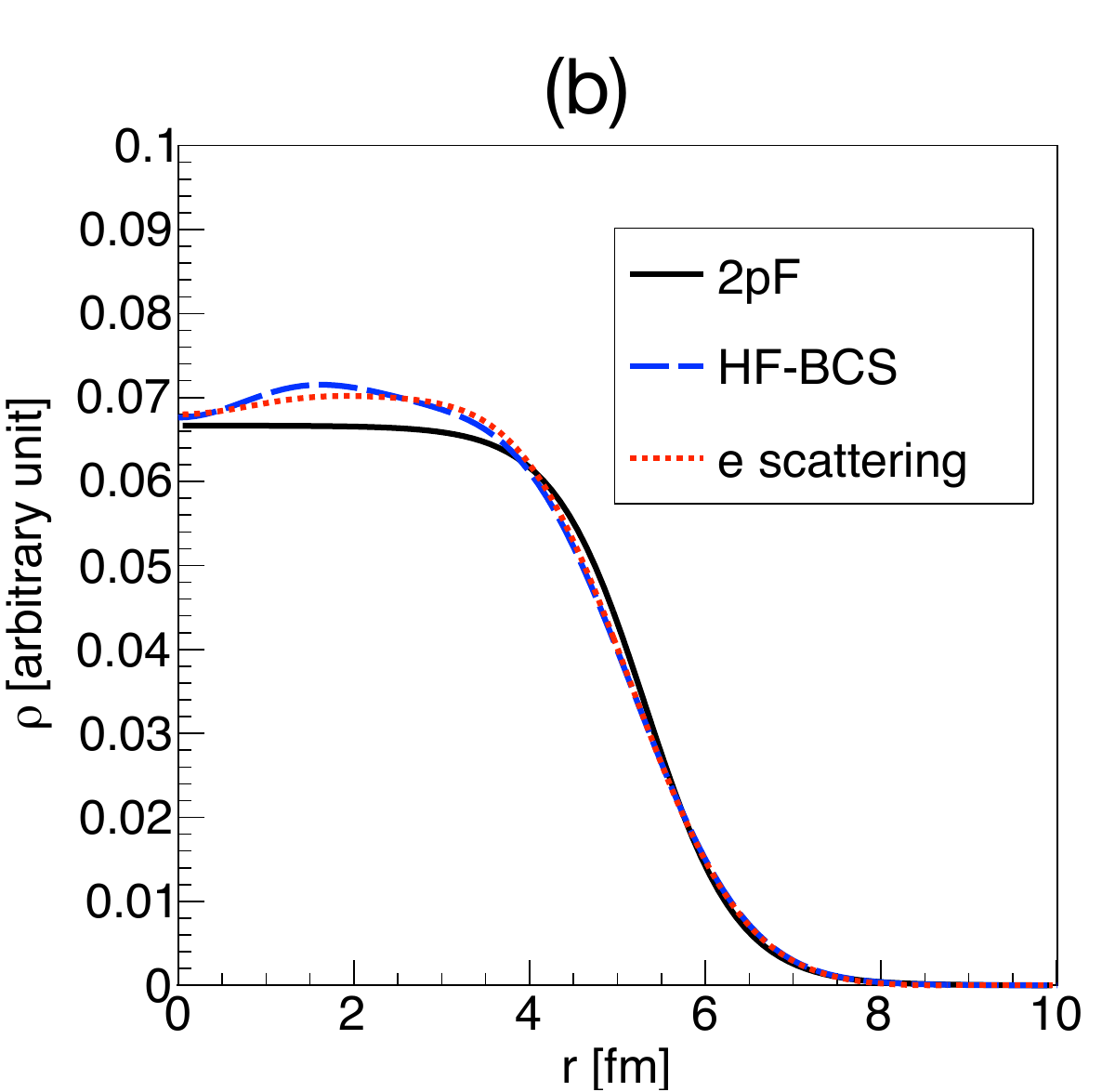}
  \caption{ (a) Barrett equivalent radii for each index $k$. Five transitions of $2p$-$1s$, $3d$-$2p$, and $4d$-$2p$ are plotted. Note that two transitions of $2p$-$1s$ are overlapped each other. (b) Nuclear charge distribution with different functional forms for $\mathrm{^{108}Pd}$. The black solid lines are 2pF distribution, the blue dashed lines are the theoretical calculation from HF-BCS and the red dotted lines are from the electron scattering experiment. 
  }
  \label{fig:barrett}
\end{figure}

The physical interpretation of the Barrett model becomes clear comparing it with the ordinary nuclear moments.
In Barrett's approach, the equivalent radius $R_{k\alpha_B}$ can be determined for each $k$ and $\alpha_B$ while the rms radius is not obtained.
Considering $\alpha_B$ shown in Table~\ref{tab:xray} is almost constant around $0.1$ for all transitions, let us consider the case of $\alpha_B = 0$.
Under this simplification, the Barrett moment (Eq.~(\ref{eq:barrettmoment})) is now a $k$-th nuclear moment with a non-integer order index.
The index $k$ increases from $2$ to $6$ as the transitions go higher, and the $k$-th nuclear moments are estimated for each transition.
Thus, the measurement of the higher transitions provides information on the higher nuclear moments. 
Furthermore, the simplified Barrett moments provide a clear view of the benefits of the higher transitions to determine the charge parameters of the 2pF function discussed in Sect.~\ref{sec_radius}.
The series of the nuclear moments given by Barrett's approach corresponds to the difference in the gradient of the 1$\sigma$ region in Fig.~\ref{fig:fermi2d}-(a).
Therefore, two $2p$-$1s$ transitions with similar $k$ give similar gradients in Fig.~\ref{fig:fermi2d}-(a) and the $3d$-$2p$ transitions with $k \neq 2$ help to limit the $c$-$t$ space of the 2pF distribution.

The $k$-dependence of the equivalent radius can be used to discuss the general functional forms of the charge distribution since the discussion of the equivalent radii is now not limited to 2pF.
Figure~\ref{fig:barrett}-(a) shows the measured equivalent radii for ${\rm ^{108}Pd}$ for the $2p$-$1s$ and $3d$-$2p$ transitions as a function of $k$.
Hereafter, we assume $\alpha_B$ is constant among the transitions and use the averaged value of two $2p$-$1s$ transitions as $\alpha_B$.
Once a functional form of the charge distribution is given, one can calculate the equivalent radii corresponding to the charge distribution for any $k$ and draw the curve in this $k$-$R_{k\alpha}$ space as shown in Fig.~\ref{fig:barrett}-(a).
The lines drawn in Fig.~\ref{fig:barrett}-(a) represent the equivalent radius lines corresponding to the three charge distributions shown in Fig.~\ref{fig:barrett}-(b).
The black solid lines in the figures represent 2pF with the Fermi parameters in Table~\ref{tab:xray}. 
Since this 2pF is considered to be the non-perturbative distribution $\rho_0 (r)$ in this treatment, this agreement just indicates self-consistency in the present analysis.
The blue dashed lines are a Hartree-Fock plus BCS calculation (HF-BCS) in 3D coordinate space with the SkM* effective interaction and the monopole pairing interaction~\cite{InPACS, Ebata2014,Ebata2017,Bartel1982}, and the red dotted lines represent the electron scattering experiment~\cite{Laan1986}.
All of the charge distributions fulfill the normalization of $\int \rho (r) 4\pi r^2 \mathrm{d}r = Z$.

Using Barrett's approach, one can compare the muonic X-ray transition energies with the given charge distribution in the $k$-$R_{k\alpha_B}$ space.
The blue dashed lines in Fig.~\ref{fig:barrett} show the theoretical calculation using HF-BCS with 3D Skyrme force.
Since the original reference provides the density distribution of the point proton, the charge distribution is calculated by folding with the Gaussian proton charge distribution~\cite{Kurasawa2019}.
The theoretical distribution gives a rms charge radius of 4.55~fm, which is consistent with that the obtained value of 4.537$_{-6}^{+27}$~fm in the present study.
This consistency in the rms radius corresponds to the agreement in the equivalent radii for the $2p$-$1s$ transitions in Fig.~\ref{fig:barrett}-(a) since the Barrett moments at $k \sim 2$ are very similar to the second-order nuclear moment, namely the $ \langle r^2 \rangle = (\mathrm{rms})^2 $.
On the contrary, the theoretical calculation overestimates the Barrett moments at the $3d$-$2p$ transitions at $k \sim 3.7$.
It indicates that the theoretical calculation fails to reproduce the higher moment despite the agreement on the rms radius.
The present example demonstrates that the measurement of the higher transitions provides a further constraint to the theoretical calculations.




The red dotted line in Fig.~\ref{fig:barrett}-(b) is the charge distribution obtained from the electron scattering for $^{108}\mathrm{Pd} $~\cite{Vries1987, Laan1986}.
As shown in Fig.~\ref{fig:barrett}-(a), the equivalent radii obtained from the present X-ray measurement agree with those from the electron scattering for the $3d$-$2p$ transitions at $k \sim 3.7$.
On the other hand, our equivalent radii are larger than those obtained from the electron scattering at $k \sim 2$.
It reflects the fact that the rms radius obtained from muonic X-ray spectroscopy (4.537$_{-6}^{+27}$~fm) is slightly larger than 4.524(10)~fm from the electron scattering.
Note that two independent measurements provide almost the same accuracy for the rms radius.
As discussed in Sect.~\ref{sec_radius}, the uncertainty of the rms radius is dominated by that of the $3d$-$2p$ transition energies.
The uncertainty of the equivalent radii for the $2p$-$1s$ transitions is thus smaller than that of the rms radius.
Since the equivalent radii are linked to the individual muonic transitions while the rms radius is deduced by combining several transitions in the present analysis, the $k$-$R_{k\alpha_B}$ space provides a direct comparison among the different experimental data.
Although the quantitative discussion is difficult because the uncertainty of the distribution of the electron scattering is not given in the reference, the discrepancy is probably due to the difference in the sensitivity to the charge distribution between the two methods.
The scattering experiment usually quotes larger uncertainty for the inner part of the distribution because the experimental uncertainty is larger for the larger momentum transfer. 
On the other hand, the muonic transition energies have the smallest uncertainty for the $k\sim 2$, namely the $2p$-$1s$ transitions, and thus it is rather sensitive to the center of the distribution.
Therefore, two experimental methods, namely the muonic X-ray spectroscopy and the electron scattering, are complementary to discuss the charge distribution.


We have demonstrated that the muonic transition energies constrain the nuclear charge distribution through the $k$-$R_{k\alpha_B}$ plot.
In the present work, the systematic comparison of the Barrett radii along the index $k$ is limited by the experimental uncertainty of the $3d$-$2p$ transition energy, which is similar to the derivation of the rms radius discussed in Sect.~\ref{sec_radius}.
For further constraints on the charge distribution from the muonic X-ray spectroscopy, a comprehensive observation of the X-ray series with a variety of $k$ wider than the present work is required. 
As $k$ represents the order index of the Barret moments, the higher series of the muonic X-ray, such as $4f$-$3d$ at $k\sim6$, restricts a higher moment, namely the outer region of the charge distribution.
Furthermore, the transitions with smaller $k$, such as $2s$-$2p$ at $k\sim 1.5$, for example, are particularly of interest.
The transition energies with small $k$ constrain the inner part of the distribution, which is difficult to investigate with the electron scattering. 
The inner part of the nuclear charge distribution will be discussed with a future high-statistics and high-resolution measurements using the $k$-$R_{k\alpha_B}$ plot.

\section{Conclusion}\label{sec_conclusion}

The muonic X-ray spectroscopy experiment was performed at MuSIC-M1 beamline at Research Center for Nuclear Physics (RCNP), Osaka University. 
The muonic X-ray energies were measured up to the $4f$-$3d$ transitions for stable palladium isotopes with $A= 104$, $105$, $106$, $108$ and $110$.
The analysis method to deduce the root-mean-square (rms) charge radius from the muonic transition energies is proposed.  
By combining the $2p$-$1s$ and $3d$-$2p$ transition energies, two parameters in the two-parameter Fermi (2pF) distribution are simultaneously deduced and the rms radius is obtained with the experimental uncertainty. 
The charge distribution of the nucleus is discussed by employing the Barrett model.
The $k$-$R_{k\alpha_B}$ plot provides a direct comparison between the muonic transition energies and the charge distribution resulting from the electron scattering and theoretical calculations.

\section*{Acknowledgements}

The authors would like to acknowledge the accelerator team at RCNP, Osaka University for providing the stable proton primary beam during the experiment. 
Fruitful discussions with Dr.~T.~Naito and Dr.~F.~Minato are greatly appreciated. 
This work is funded by the ImPACT Program of the Council for Science, Technology, and Innovation (Cabinet Office, Government of Japan), and partially supported by JSPS KAKENHI Grant Number JP18J10554.

\bibliography{PRC_music2016}

\end{document}